\documentclass[12pt]{iopart}

\def\vereq#1#2{\lower3pt\vbox{\baselineskip1.5pt \lineskip1.5pt
\ialign{$#1\hfill##\hfil$\crcr#2\crcr\sim\crcr}}}

\newcommand{\gtsim}{\mathrel{\mathpalette\vereq>}}

\begin{document}

\jl{01}
\date{\today}
\centerline{\submitted\hfill\today}

\title{Decimation and harmonic inversion of periodic orbit signals}
\author{J Main\dag, P A Dando\ddag, D\v z Belki\'c$\ddag$ and H S Taylor\ddag}
\address{\dag\ Institut f\"ur Theoretische Physik und Synergetik,
         Universit\"at Stuttgart,\\ D-70550 Stuttgart, Germany}
\address{\ddag\ Department of Chemistry, University of Southern California,\\
         Los Angeles, CA 90089, USA}

\begin{abstract}
We present and compare three generically applicable signal processing 
methods for periodic orbit quantization via 
harmonic inversion of semiclassical recurrence functions.
In a first step of each method, a band-limited decimated periodic orbit 
signal is obtained by analytical frequency windowing of the periodic orbit sum.
In a second step, the frequencies and amplitudes of the decimated
signal are determined by either Decimated Linear Predictor, 
Decimated Pad\'e Approximant, or Decimated Signal Diagonalization.
These techniques, which would have been numerically unstable without the
windowing,  provide numerically more accurate semiclassical spectra 
than does the filter-diagonalization method.
\end{abstract}


\section{Introduction}
\label{intro}
The semiclassical quantization of systems with an underlying chaotic
classical dynamics is a nontrivial problem for the reason that Gutzwiller's
trace formula \cite{Gut67,Gut90} does not usually converge in those
regions where the eigenenergies or resonances are located.
Various techniques have been developed to circumvent the convergence 
problem of periodic orbit theory.
Examples are the cycle expansion technique \cite{Cvi89}, the Riemann-Siegel 
type formula and pseudo-orbit expansions \cite{Ber90}, surface of section
techniques \cite{Bog92}, and a quantization rule based on a semiclassical
approximation to the spectral staircase \cite{Aur92}.
These techniques have proven to be very efficient for systems with 
special properties, {\it e.g.\/}, the cycle expansion for hyperbolic systems 
with an existing symbolic dynamics, while the other mentioned methods 
have been used for the calculation of bound spectra.

Recently, an alternative method based upon Filter-Diagonalization (FD) 
has been introduced for the analytic continuation of the semiclassical 
trace formula \cite{Mai97a,Mai98a}.
The FD method requires knowledge of the periodic orbits up to a given maximum 
period (classical action), which depends on the mean density of states.
The same holds true for the three methods presented in this paper.
The semiclassical eigenenergies or resonances are obtained by 
{\em harmonic inversion\/} of the periodic orbit recurrence signal.
The FD method can be generally applied to both open and bound systems and has 
also proven powerful, {\it e.g.\/}, for the calculation of semiclassical
transition matrix elements \cite{Mai99a} and the quantization of systems
with mixed regular-chaotic phase space \cite{Mai99b}.
For a review on periodic orbit quantization by harmonic inversion see 
\cite{Mai99c}.

In this paper the techniques for the harmonic inversion of periodic orbit
signals are further developed.
The semiclassical signal, in action or time, corresponds to a ``spectrum''
or response in the frequency domain
that is composed of a huge, in principle infinite, number of frequencies.
To extract these frequencies and their corresponding amplitudes
is a nontrivial task.
In the previous work \cite{Mai97a,Mai98a,Mai99c} the periodic orbit signal
has been harmonically inverted by means of FD
\cite{Wal95,Man97a,Man97b}, which is designed for the analysis of
time signals given on an equidistant grid.
The periodic orbit recurrence signal is given as a sum over usually unevenly
spaced $\delta$ functions. 
A smooth signal, from which evenly spaced values can be read off, is obtained 
by a convolution of this sum with, {\it e.g.\/}, a narrow Gaussian function.
The disadvantages of this approach are twofold.
Firstly, FD acts on this signal more or less like a ``black box''
and, as such, does not lend itself to a detailed understanding of 
semiclassical periodic orbit quantization.
Secondly, the smoothed semiclassical signal usually consists of a huge 
number of data points.
The handling of such large data sets, together with the smoothing, may lead 
to significant numerical errors in some of the results for the semiclassical 
eigenenergies and resonances.

Here, we propose three alternative methods for the harmonic inversion of
the periodic orbit recurrence signal that avoid these problems.
In a first step we create a shortened signal which is constructed 
from the original signal and designed to be correct only in a window, 
{\it i.e.\/}, a short frequency range of the total band width.
Because the original signal is given as a periodic orbit sum of 
$\delta$ functions, this ``filtering'' can be performed analytically 
resulting in a decimated periodic orbit signal with a relatively small number 
of equidistant grid points.
In a second step the frequencies and amplitudes of the decimated signal
are determined from a set of nonlinear equations.
To solve the nonlinear system, we introduce three different processing methods:
Decimated Linear Predictor (DLP), 
Decimated Pad\'e Approximant (DPA), and 
Decimated Signal Diagonalization (DSD).
The standard and well-known Linear Predictor (LP) and Pad\'e Approximant (PA)
would not have yielded numerically stable solutions if the signal had not 
first been decimated by the windowing (filtering) procedure.
Furthermore, this separation of the harmonic inversion procedure into 
various steps may lead to a clearer picture of the periodic orbit 
quantization method.
Numerical examples will demonstrate that the techniques proposed in this 
paper also provide more accurate results than previous applications of FD.

The paper is organized as follows.
In \sref{po_quant:sec} we briefly review the general idea of periodic orbit 
quantization by harmonic inversion.
In \sref{analyt_dec:sec} we construct the band-limited decimated periodic
orbit signal which is analyzed in \sref{hi:sec} with the help of three
different methods, {\it viz.\/}\ DLP, DPA and DSD.
In \sref{results:sec} we present and compare results for the three-disk 
scattering system as a physical example and the zeros of the Riemann zeta 
function as a mathematical model for periodic orbit quantization.
Some concluding remarks are given in \sref{concl:sec}.

\section{Periodic orbit quantization by harmonic inversion}
\label{po_quant:sec}
In order to understand the following, a brief recapitulation of  
the basic ideas of periodic orbit quantization by 
harmonic inversion is necessary.
For further details see \cite{Mai99c}.

Following Gutzwiller \cite{Gut67,Gut90} the semiclassical response function
for chaotic systems is given by
\begin{equation}
 g^{\rm sc}(E) = g^{\rm sc}_0(E)
   + \sum_{\rm po} {\cal A}_{\rm po} e^{iS_{\rm po}} \; ,
\label{gE_sc}
\end{equation}
where $g^{\rm sc}_0(E)$ is a smooth function and $S_{\rm po}$ and 
${\cal A}_{\rm po}$ are the classical actions and weights (including phase 
information given by the Maslov index) of the periodic orbit contributions.
Equation~\eref{gE_sc} is also valid for integrable systems when the 
periodic orbit
quantities are calculated not with Gutzwiller's trace formula, but with the 
Berry-Tabor formula \cite{Ber76} for periodic orbits on rational tori.
The eigenenergies and resonances are the poles of the response function.
Unfortunately, the semiclassical approximation \eref{gE_sc} does
not converge in the region of the poles and hence the problem is the analytic 
continuation of $g^{\rm sc}(E)$ to this region.

As done previously \cite{Mai97a,Mai98a,Mai99c}, we will also make the (weak) 
assumption 
that the classical system has a scaling property, {\it i.e.\/}, the shape 
of periodic orbits does not depend on the scaling parameter, $w$, and the 
classical action scales as 
\begin{equation}
 S_{\rm po} = ws_{\rm po} \; .
\label{S_po}
\end{equation}
In scaling systems, the fluctuating part of the semiclassical 
response function,
\begin{equation}
 g^{\rm sc}(w) = \sum_{\rm po} {\cal A}_{\rm po} e^{iws_{\rm po}} \; ,
\label{g_sc}
\end{equation}
can be Fourier transformed readily to obtain the semiclassical trace of 
the propagator
\begin{equation}
   C^{\rm sc}(s) 
 = {1 \over 2\pi} \int_{-\infty}^{+\infty} g^{\rm sc}(w) e^{-isw} dw
 = \sum_{\rm po} {\cal A}_{\rm po} \delta\left(s-s_{\rm po}\right) \; .
\label{C_sc}
\end{equation}
The signal $C^{\rm sc}(s)$ has $\delta$-spikes at the positions of the 
classical periods (scaled actions) $s=s_{\rm po}$ of periodic orbits and 
with peak heights (recurrence strengths) ${\cal A}_{\rm po}$, {\it i.e.\/}, 
$C^{\rm sc}(s)$ is Gutzwiller's periodic orbit recurrence function.
Consider now the quantum mechanical counterparts of $g^{\rm sc}(w)$ and
$C^{\rm sc}(s)$ taken as the sums over the poles $w_k$ of the Green's 
function,
\begin{equation}
 g^{\rm qm}(w) = \sum_k {d_k \over w-w_k+i\epsilon} \; ,
\label{g_qm}
\end{equation}
\begin{equation}
   C^{\rm qm}(s)
 = {1\over 2\pi} \int_{-\infty}^{+\infty} g^{\rm qm}(w) e^{-isw} dw
 = -i\sum_k d_k e^{-i w_k s} \; ,
\label{C_qm}
\end{equation}
with $d_k$ being the residues associated with the eigenvalues.
In the case under study, {\it i.e.\/}, density of state spectra,
the $d_k$ are the multiplicities of eigenvalues and should be equal
to 1 for non-degenerate states.
Semiclassical eigenenergies $w_k$ and residues $d_k$ can now,
in principle, be obtained by adjusting the semiclassical signal,
eq.~\eref{C_sc}, to the functional form of the quantum signal, 
eq.~\eref{C_qm}, with the $\{w_k,d_k\}$ being free generally 
complex frequencies and amplitudes.
This procedure is known as ``harmonic inversion''.
The numerical procedure of harmonic inversion is a nontrivial task,
especially if the number of frequencies in the signal is large 
({\it e.g.\/}, more than a thousand) or even infinite as is usually the case 
for periodic orbit quantization.
Note that the conventional way to perform the spectral analysis, {\it i.e.\/},
the Fourier transform of eq.~\eref{C_sc} will bring us back to analyzing
the non-convergent response function $g^{\rm sc}(w)$ in eq.~\eref{g_sc}.
The periodic orbit signal \eref{C_sc} can be harmonically inverted by
application of FD \cite{Wal95,Man97a,Man97b},
which allows one to calculate a finite and relatively small set of
frequencies and amplitudes in a given frequency window.
The usual implementation of FD requires knowledge of the signal on an 
equidistant grid.
The signal \eref{C_sc} is not a continuous function.
However, a smooth signal can be obtained by a convolution of $C^{\rm sc}(s)$ 
with, {\it e.g.\/}, a Gaussian function,
\begin{equation}
   C_\sigma^{\rm sc}(s) 
 = {1\over \sqrt{2\pi}\sigma}
   \sum_{\rm po} {\cal A}_{\rm po} e^{-(s-s_{\rm po})^2/2\sigma^2} \; .
\label{C_sc_sigma}
\end{equation}
As can easily be seen, the convolution results in a damping of the 
amplitudes, $d_k\to d_k^{(\sigma)}=d_k\exp{(-w_k^2\sigma^2/2)}$.
The width $\sigma$ of the Gaussian function should be chosen sufficiently 
small to avoid an overly strong damping of amplitudes.
To properly sample each Gaussian a dense grid with steps
$\Delta s\approx\sigma/3$ is required.
Therefore, the signal \eref{C_sc_sigma} analyzed by FD
usually consists of a large number of data points.
The numerical treatment of this large data set may suffer from rounding
errors and loss of accuracy.
Additionally, the ``black box'' type procedure of harmonic inversion by
FD, which intertwines windowing and processing,
does not provide any opportunity to seek a deeper 
understanding of semiclassical periodic orbit quantization.
It is therefore desirable to separate the harmonic inversion procedure into
two sequential steps:
Firstly, the filtering procedure that does not require smoothing
and, secondly, a procedure for extracting  the frequencies and amplitudes.
In \sref{analyt_dec:sec} we will construct, by analytic filtering, a
band-limited signal which consists of a relatively small number of frequencies.
In \sref{hi:sec} we will present methods to extract the frequencies and 
amplitudes of such band-limited decimated signals.

\section{Construction of band-limited decimated signals by analytical 
filtering}
\label{analyt_dec:sec}
Consider a signal of a presumably large length $N$.
We split the corresponding Fourier spectrum, which is also of length $N$, 
into $M$ frequency intervals.
In general, a frequency filter can be applied to a given signal by
application of the Fourier transform \cite{Zol91,Bel99}.
Specifically, the signal, in time or action, is first transformed to the 
frequency domain, {\it e.g.\/}, by application of the fast Fourier transform 
(FFT) or by using the closed-form expression, if available, for the Fourier 
integral.
The transformed signal, which is essentially a low-resolution spectrum,
is multiplied with a frequency filter function $f(w)$ localized around 
a central frequency, $w_0$, and zeroed out everywhere else outside the 
selected window.
This leads to a band-limited Fourier spectrum.
The frequency filter $f(w)$ can be rather general; typical examples are 
a rectangular window or a Gaussian function.
The resulting filtered or band-limited spectrum is then shifted by $-w_0$,
relocated symmetrically around the frequency origin, $w=0$, and transformed 
back to the time domain by the application of the inverse FFT to produce a 
band-limited signal valid only in the window defined by the filter function.
Such a band-limited signal still has the same original length $N$.
It is at this step that we apply decimation which amounts to an enhancement
of the sampling time by a factor of $M$ thus leading finally to a band-limited
decimated signal of length $[N/M]$, where $[u]$ denotes the integer part of 
$u$.
In other words, since the band-width of the band-limited decimated signal 
is $M$ times smaller than that of the original signal, the sampling rate, 
or dwell time, between signal samplings can now be $M$ times larger.
Hence, the filtered or band-limited signal can be reduced by retaining only
those signal points for which the time or action indices are, say,
$1, M+1, 2M+1,\ldots$.
This technique is known as ``beamspacing'' \cite{Zol91} or ``decimation'' 
\cite{Bel99} of band-limited signals.
The flexibility in the
choice of window size can ensure a numerically stable implementation of the
processing methods presented below.

The special form of the periodic orbit signal \eref{C_sc} as a sum of
$\delta$ functions allows for an even simpler procedure, {\it viz.\/}\
analytical filtering.
In the following we will apply a rectangular filter, {\it i.e.\/}, $f(w)=1$ 
for frequencies $w \in [w_0-\Delta w,w_0+\Delta w]$, and $f(w)=0$ outside 
the window.
Generalization to other types of frequency filters is straightforward.
Starting from the semiclassical response function (spectrum)
$g^{\rm sc}(w)$ in eq.~\eref{g_sc}, which is itself a Fourier transform 
of the ``signal'' \eref{C_sc}, and using a rectangular window we obtain,
after evaluating the ``second'' Fourier transform, the band-limited (bl) 
periodic orbit signal,
\begin{eqnarray}
     C^{\rm sc}_{\rm bl}(s)
 &=& {1\over 2\pi} \int_{w_0-\Delta w}^{w_0+\Delta w}
     g^{\rm sc}(w) e^{-is(w-w_0)} dw \nonumber \\
 &=& {1\over 2\pi} \sum_{\rm po} {\cal A}_{\rm po}
     \int_{w_0-\Delta w}^{w_0+\Delta w}
     e^{isw_0-i(s-s_{\rm po})w} dw \nonumber \\
 &=& \sum_{\rm po} {\cal A}_{\rm po} {\sin{[(s-s_{\rm po})\Delta w]}\over
     \pi(s-s_{\rm po})} e^{is_{\rm po}w_0} \; .
\label{C_sc_bl}
\end{eqnarray}
The introduction of $w_0$ into the arguments of the exponential 
functions in \eref{C_sc_bl}, causes a shift of frequencies by $-w_0$ 
in the frequency domain.
Note that $C^{\rm sc}_{\rm bl}(s)$ is a smooth function and can be easily
evaluated on an arbitrary grid of points $s_n<s_{\rm max}$ provided the
periodic orbit data are known for the set of orbits with classical action
$s_{\rm po}<s_{\rm max}$.

Applying now the same filter used for the semiclassical periodic orbit 
signal to the quantum one, we obtain the band-limited quantum signal
\begin{eqnarray}
     C^{\rm qm}_{\rm bl}(s)
 &=& {1\over 2\pi} \int_{w_0-\Delta w}^{w_0+\Delta w}
     g^{\rm qm}(w) e^{-is(w-w_0)} dw  \nonumber \\
 &=& -i \sum_{k=1}^K d_k e^{-i(w_k-w_0)s} \; , \quad |w_k-w_0| < \Delta w \; .
\label{C_qm_bl}
\end{eqnarray}
In contrast to the signal $C^{\rm qm}(s)$ in eq.~\eref{C_qm}, the band-limited
quantum signal consists of a {\em finite} number of frequencies $w_k$, 
$k=1,\dots,K$, where $K$ can be of the order of $\sim$~(50-200) for 
an appropriately chosen frequency window, $\Delta w$.
The problem of adjusting the band-limited semiclassical signal in
eq.~\eref{C_sc_bl} to its quantum mechanical analogue in eq.~\eref{C_qm_bl}
can now be written as a set of $2K$ nonlinear equations
\begin{equation}
   C^{\rm sc}_{\rm bl}(n\tau) \equiv c_n
 = -i \sum_{k=1}^K d_k e^{-iw'_kn\tau} \; , \quad n=0,1,\dots,2K-1 \; ,
\label{C_bld}
\end{equation}
for the $2K$ unknown variables, {\it viz.\/}\ the shifted frequencies, 
$w'_k\equiv w_k-w_0$, and amplitudes, $d_k$.
The signal now becomes ``short'' (decimated) as it can be evaluated on 
an equidistant grid, $s=n\tau$, with step width $\tau\equiv\pi/\Delta w$.
It is important to note that the number of signal points $c_n$ in 
eq.~\eref{C_bld} is usually much smaller than a reasonable discretization
of the signal $C_\sigma^{\rm sc}(s)$ in eq.~\eref{C_sc_sigma}, which is
the starting point for harmonic inversion by FD.
Therefore, the discrete signal points $c_n\equiv C^{\rm sc}_{\rm bl}(n\tau)$ 
are called the ``band-limited decimated'' periodic orbit signal.
Methods to solve the nonlinear system, eq.~\eref{C_bld}, are discussed in
\sref{hi:sec} below.

It should also be noted that the analytical filtering in eq.~\eref{C_sc_bl}
is not restricted to periodic orbit signals, but can be applied,
in general, to any signal given as a sum of $\delta$ functions.
An example is the high resolution analysis of quantum spectra 
\cite{Mai97b,Mai99c}, where the density of states reads
$\varrho(E)=\sum_n\delta(E-E_n)$.

\section{Harmonic inversion of decimated signals}
\label{hi:sec}
In this section we want to solve the nonlinear set of equations
\begin{equation}
 c_n = \sum_{k=1}^K d_k z_k^n \; , \quad n=0,1,\dots,2K-1 \; ,
\label{c_n:eq}
\end{equation}
where $z_k\equiv\exp{(-iw'_k\tau)}$ and $d_k$ are generally complex 
variational parameters.
For notational simplicity we have absorbed the factor of $-i$ on the 
right-hand side of eq.~\eref{C_bld} into the $d_k$'s with the understanding 
that this should be corrected for at the end of the calculation.
We assume that the number of frequencies in the signal is relatively small
($K\sim 50$ to $200$).
Although the system of nonlinear equations is, in general, still 
ill-conditioned, frequency filtering reduces the number of signal points, 
and hence the number of equations.
Several numerical techniques, that otherwise would be numerically unstable,
can now be applied successfully.
In the following we introduce three different methods, {\it viz.\/}\
Decimated Linear Predictor (DLP), 
Decimated Pad\'e Approximant (DPA), and
Decimated Signal Diagonalization (DSD).

\subsection{Decimated Linear Predictor}
The problem of solving eq.~\eref{c_n:eq} has already been addressed 
in the 18th century by Baron de Prony, who converted the nonlinear set 
of equations \eref{c_n:eq} to a linear algebra problem.
Today this method is known as Linear Predictor (LP).
Our method, called Decimated Linear Predictor (DLP), strictly applies the 
procedure of LP except with one essential difference;
the original signal $C^{\rm sc}(s)$ is replaced by its band-limited decimated 
counterpart $c_n \equiv C^{\rm sc}_{\rm bl}(n\tau)$.

Equation~\eref{c_n:eq} can be written in matrix form for the signal points
$c_{n+1}$ to $c_{n+K}$,
\begin{equation}
   \left(\begin{array}{c} c_{n+1}\\ \vdots \\ c_{n+K} \end{array} \right)
 = \left(\begin{array}{ccc}
      z_1^{n+1} & \cdots & z_K^{n+1} \\
      \vdots & & \vdots \\
      z_1^{n+K} & \cdots & z_K^{n+K}
   \end{array} \right)
   \left(\begin{array}{c} d_1 \\ \vdots \\ d_K \end{array} \right) \; .
\label{c_n_matrix}
\end{equation}
From the matrix representation \eref{c_n_matrix} it follows that
\begin{equation}
c_n = \left( z_1^n \cdots z_K^n \right)
      \left(\begin{array}{ccc}
          z_1^{n+1} & \cdots & z_K^{n+1} \\
          \vdots & & \vdots \\
          z_1^{n+K} & \cdots & z_K^{n+K}
      \end{array} \right)^{-1}
      \left(\begin{array}{c} c_{n+1} \\ \vdots \\ c_{n+K} \end{array} \right)
    = \sum_{k=1}^K a_k c_{n+k} \; ,
\label{a_k_def}
\end{equation}
which means that every signal point $c_n$ can be ``predicted'' by a linear
combination of the $K$ subsequent points with a fixed set of coefficients 
$a_k$, $k=1,\dots,K$.
The first step of the LP method is to calculate these coefficients.
Writing eq.~\eref{a_k_def} in matrix form with $n=0,\dots,K-1$, we obtain 
the coefficients $a_k$ as solution of the linear set of equations,
\begin{equation}
   \left(\begin{array}{ccc}
       c_{1} & \cdots & c_{K} \\
       \vdots & & \vdots \\
       c_{K} & \cdots & c_{2K-1}
   \end{array} \right)
   \left(\begin{array}{c} a_1 \\ \vdots \\ a_K \end{array} \right)
 = \left(\begin{array}{c} c_0 \\ \vdots \\ c_{K-1} \end{array} \right) \; .
\label{lin_eq1}
\end{equation}
The second step is the determination of the parameters $z_k$ in 
eq.~\eref{c_n:eq}.
Using eqs.~\eref{a_k_def} and \eref{c_n:eq} we obtain
\begin{equation}
   c_n=\sum_{k=1}^K a_k c_{n+k}
 = \sum_{l=1}^K \sum_{k=1}^K a_k d_l z_l^{n+k} \; ,
\end{equation}
and thus
\begin{equation}
 \sum_{k=1}^K \left[ \sum_{l=1}^K a_l z_k^{n+l}-z_k^n \right] d_k = 0 \; .
\label{poly1:eq}
\end{equation}
Equation~\eref{poly1:eq} is satisfied for arbitrary sets of amplitudes $d_k$
when $z_k$ is a zero of the polynomial
\begin{equation}
 \sum_{l=1}^K a_l z^l - 1 = 0 \; .
\label{poly2:eq}
\end{equation}
The parameters $z_k=\exp{(-iw'_k\tau)}$ and thus the frequencies
\begin{equation}
 w'_k = {i\over\tau} \log(z_k)
\label{wk:eq}
\end{equation}
are therefore obtained by searching for the zeros of the polynomial 
in eq.~\eref{poly2:eq}.
Note that this is the only nonlinear step of the algorithm and numerical
routines for finding the roots of polynomials are well established.
In the third and final step, the amplitudes $d_k$ are obtained from the
linear set of equations
\begin{equation}
c_n = \sum_{k=1}^K d_k z_k^n \; , \quad n = 0, \dots, K-1 \; .
\label{lin_eq2}
\end{equation}
To summarize, the LP method reduces the {\em nonlinear\/} set of 
equations \eref{c_n:eq} for the variational parameters $\{z_k,d_k\}$ to 
two well-known problems, {\it i.e.\/}, the solution of two {\em linear\/} 
sets of equations \eref{lin_eq1} and \eref{lin_eq2} and the root search of a
polynomial, eq.~\eref{poly2:eq}, which is a nonlinear but familiar problem.
The matrices in eqs.~\eref{lin_eq1} and \eref{lin_eq2} are a Toeplitz and 
Vandermonde matrix, respectively, and special algorithms are known for
the fast solution of such linear systems \cite{NumRec}.
However, when the matrices are ill-conditioned, conventional $LU$
decomposition of the matrices is numerically more stable, and, furthermore,
an iterative improvement of the solution can significantly reduce errors 
arising from numerical round-off.
The roots of polynomials can be found, in principle, by application
of Laguerre's method \cite{NumRec}.
However, it turns out that an alternative method, {\it i.e.\/}, the 
diagonalization of the Hessenberg matrix 
\begin{equation}
 {\bf A} =
 \left(\begin{array}{ccccc}
  -{a_{K-1}\over a_K} & -{a_{K-2}\over a_K} & \cdots &
  -{a_{1}\over a_K} & -{a_{0}\over a_K} \\
  1 & 0 & \cdots & 0 & 0 \\
  0 & 1 & \cdots & 0 & 0 \\
  \vdots & & & & \vdots \\
  0 & 0 & \cdots & 1 & 0 
 \end{array} \right) \quad ,
\label{Hesse:eq}
\end{equation}
for which the characteristic polynomial $P(z)=\det[{\bf A}-z{\bf I}]=0$ 
is given by eq.~\eref{poly2:eq} (with $a_0=-1$), is a numerically more 
robust technique for finding the roots of high degree ($K \gtsim 60$) 
polynomials \cite{NumRec}.

\subsection{Decimated Pad\'e Approximant}
As an alternative method for solving the nonlinear system \eref{c_n:eq}
we now propose to apply the method of Decimated Pad\'e Approximants (DPA).
This is the standard Pad\'e Approximant (PA) but applied to our band-limited
decimated signal $c_n$.
Let us assume for the moment that the signal points $c_n$ are known up
to infinity, $n=0,1,\dots\infty$.
Interpreting the $c_n$'s as the coefficients of a Maclaurin series in the
variable $z^{-1}$, we can then define the function 
$g(z)=\sum_{n=0}^\infty c_n z^{-n}$.
With eq.~\eref{c_n:eq} and the sum rule for geometric series we obtain
\begin{equation}
   g(z) \equiv \sum_{n=0}^\infty c_n z^{-n}
 = \sum_{k=1}^K d_k \sum_{n=0}^\infty (z_k/z)^n
 = \sum_{k=1}^K {z d_k \over z-z_k} 
 \equiv {P_{K}(z) \over Q_K(z)} \; .
\label{g_Pade:eq}
\end{equation}
The right-hand side of eq.~\eref{g_Pade:eq} is a rational function 
with polynomials of degree $K$ in the numerator and denominator.
Evidently, the parameters $z_k=\exp{(-iw'_k\tau)}$ are the
poles of $g(z)$, {\it i.e.\/}, the zeros of the polynomial $Q_K(z)$.
The parameters $d_k$ are calculated via the residues of the last two terms 
of \eref{g_Pade:eq}.
We obtain
\begin{equation}
 d_k = {P_{K}(z_k) \over z_k Q'_K(z_k)} \; ,
\label{dk_pade:eq}
\end{equation}
with the prime indicating the derivative $d/dz$.
Of course, the assumption that the coefficients $c_n$ are known up to
infinity is not fulfilled  and, therefore, the sum on the left-hand side of 
eq.~\eref{g_Pade:eq} cannot be evaluated in practice.
However, the convergence of the sum can be accelerated by application of DPA.
Indeed, with DPA, knowledge of $2K$ signal points $c_0,\dots,c_{2K-1}$ is 
sufficient for the calculation of the coefficients of the two polynomials
\begin{equation}
 P_{K}(z) = \sum_{k=1}^{K} b_k z^k  \mbox{~~and~~}
 Q_K(z) = \sum_{k=1}^K a_k z^k - 1 \; .
\end{equation}
The coefficients $a_k$, $k=1,\dots,K$ are obtained as solutions of the 
linear set of equations
\begin{equation*}
 c_n = \sum_{k=1}^K a_k c_{n+k} \; ,  \quad n = 0, \dots, K-1 \; ,
\end{equation*}
which is identical to eqs.~\eref{a_k_def} and \eref{lin_eq1} for DLP.
Once the $a$'s are known, the coefficients $b_k$ are given by the 
{\em explicit}\ formula
\begin{equation}
 b_k = \sum_{m=0}^{K-k} a_{k+m} c_{m} \; , \quad k = 1, \dots , K \; .
\end{equation}
It should be noted that the different derivations of DLP and DPA 
provide the same polynomial whose zeros are the $z_k$ parameters, 
{\it i.e.\/}, the $z_k$ calculated with both methods exactly agree.
However, DLP and DPA do differ in the way the amplitudes, $d_k$, 
are calculated.
It is also important to note that DPA is applied here 
as a method for signal processing, {\it i.e.\/}, in a different context to 
that in ref.~\cite{Mai99d}, where the Pad\'e approximant is used for the 
direct summation of the periodic orbit terms in Gutzwiller's trace formula.

\subsection{ Decimated Signal Diagonalization}
In refs.~\cite{Wal95,Man97b} it has been shown how the problem of solving the nonlinear
set of equations \eref{c_n:eq} can be recast in the form of the generalized 
eigenvalue problem, 
\begin{equation}
{\bf U} \mbox{\boldmath{$B$}}_k = z_k {\bf S} \mbox{\boldmath{$B$}}_k \; .
\label{geneval}
\end{equation}
The elements of the $K\times K$ operator matrix ${\bf U}$ and
overlap matrix ${\bf S}$ depend trivially upon the $c_n$'s \cite{Man97b}:
\begin{equation}
U_{ij} = c_{i+j+1} \; ; \quad S_{ij} = c_{i+j} \; ; \quad i,j=0,\dots,K-1 \; .
\label{matels}
\end{equation}
Note that the operator matrix ${\bf U}$ is the same as in the linear system 
\eref{lin_eq1}, {\it i.e.\/} the matrix form of eq.~\eref{a_k_def} of DLP.
The matrices ${\bf U}$ and ${\bf S}$ in eq.~\eref{geneval} are complex
symmetric ({\it i.e.\/}, non-Hermitian), and the eigenvectors 
$\mbox{\boldmath{$B$}}_k$ are orthogonal with respect to the overlap 
matrix ${\bf S}$,
\begin{equation}
   \left(\mbox{\boldmath{$B$}}_k |{\bf S}| \mbox{\boldmath{$B$}}_{k'} \right)
 = N_k \delta_{kk'} \; ,
\label{Nk:eq}
\end{equation}
where the brackets define a complex symmetric inner product $(a|b)=(b|a)$,
{\it i.e.\/}, no complex conjugation of either $a$ or $b.$
The overlap matrix ${\bf S}$ is not usually positive definite
and therefore the $N_k$'s are, in general, complex normalization parameters.
An eigenvector $\mbox{\boldmath{$B$}}_k$ cannot be normalized for $N_k=0$.
The amplitudes $d_k$ in eq.~\eref{c_n:eq} are obtained from the eigenvectors
$\mbox{\boldmath{$B$}}_k$ via
\begin{equation}
 d_k = {1\over{N_k}}
 \left[ \sum_{n=0}^{K-1} c_n \mbox{\boldmath{$B$}}_{k,n} \right]^2 \; .
\label{dk_dsd:eq}
\end{equation}
The parameters $z_k$ in eq.~\eref{c_n:eq} are given as the eigenvalues of 
the generalized eigenvalue problem \eref{geneval}, and are simply related 
to the frequencies $w'_k$ in eq.~\eref{C_bld} via $z_k=\exp(-iw'_k\tau)$.

\bigskip\noindent
The three methods introduced above (DLP, DPA and DSD) look technically
quite different.
With DLP the coefficients of the characteristic polynomial \eref{poly2:eq}
and the amplitudes $d_k$ are obtained by solving two linear sets of
equations \eref{lin_eq1} and \eref{lin_eq2}.
Note that the complete set of zeros $z_k$ of eq.~\eref{poly2:eq} is required
to solve for the $d_k$ in eq.~\eref{lin_eq2}.
The DPA method is even simpler, as only one linear system, 
eq.~\eref{lin_eq1}, has to be solved to determine the coefficients
of the rational function $P_{K}(z)/Q_K(z)$.
Finding the zeros of eq.~\eref{poly2:eq} gives 
knowledge about selected parameters $z_k$, and allows one to calculate
the corresponding amplitudes $d_k$ via eq.~\eref{dk_pade:eq}.
The DSD method requires the most numerical effort, because the solution
of the generalized eigenvalue problem \eref{geneval} for both the 
eigenvalues $z_k$ and eigenvectors $\mbox{\boldmath{$B$}}_k$ is needed.

It is important to note that the three methods, in spite of their
different derivations, are mathematically equivalent and provide the same 
results for the parameters $\{z_k,d_k\}$, when the following two conditions 
are fulfilled:
the nonlinear set of equations \eref{c_n:eq} has a unique solution, when,
firstly, the matrices ${\bf U}$ and ${\bf S}$ in eq.~\eref{matels}
have a non-vanishing determinant ($\det{\bf U}\ne 0$, $\det{\bf S}\ne 0$), 
and, secondly, the parameters $z_k$ are non-degenerate 
($z_k\ne z_{k'}$ for $k\ne k'$).
These conditions guarantee the existence of a unique solution of the linear 
equations \eref{lin_eq1} and \eref{lin_eq2}, the non-singularity of the 
generalized eigenvalue problem \eref{geneval}, and the non-vanishing of both 
the derivatives $Q'_K(z_k)$ in eq.~\eref{dk_pade:eq} and the normalization 
constants $N_k$ in eqs.~\eref{Nk:eq} and \eref{dk_dsd:eq}.
Equation~\eref{c_n:eq} usually has no solution in the case of degenerate $z_k$ 
parameters.
However, degeneracies can be handled with a generalization of the ansatz 
\eref{c_n:eq} and modified equations for the calculation of the parameters.
This special case will be reported elsewhere.

While the parameters $z_k$ in eq.~\eref{c_n:eq} are usually unique, the
calculation of the frequencies $w'_k$ via eq.~\eref{wk:eq} is not unique,
because of the multivalued property of the complex logarithm.
To obtain the ``correct'' frequencies it is necessary to 
appropriately adjust the range
$\Delta w$ of the frequency filter and the step width $\tau$ of the
band-limited decimated signal \eref{C_bld}.
We recommend the following procedure.
The most convenient approach is to choose first the centre $w_0$ of the 
frequency window and the number $K$ of frequencies within that window.
Note that $K$ determines the dimension of the linear systems, and hence 
the degree of the polynomials which have to be handled numerically, and is 
therefore directly related to the computational effort required.
Frequency windows are selected to be sufficiently narrow to yield 
values for the rank between $K\approx 50$ and $K\approx 200$.
The step width for the decimated signal should be chosen as 
\begin{equation}
 \tau = {s_{\rm max}\over 2K} \; ,
\end{equation}
with $s_{\rm max}$ being the total length of the periodic orbit signal.
The relation $z_k=\exp{(-iw'_k\tau)}$ projects the frequency window
$w'\in [-\Delta w,+\Delta w]$ onto the unit circle in the complex plane
when the range of the frequency window is chosen as 
\begin{equation}
 \Delta w = {\pi\over\tau} = {2\pi K\over s_{\rm max}} \; .
\end{equation}
When calculating the complex logarithm with $\arg\log z\in [-\pi,+\pi]$, 
eq.~\eref{wk:eq} provides the ``correct'' shifted frequencies $w'_k$ and 
thus the frequencies $w_k=w_0+w'_k$.

To achieve convergence, the length $s_{\rm max}$ of the periodic orbit signal 
must be sufficiently long to ensure that the number of semiclassical 
eigenvalues within the frequency window is less than $K$.
As a consequence the harmonic inversion procedure usually provides not only
the true semiclassical eigenvalues but also some spurious resonances.
The spurious resonances are identified by low or near zero values of the
corresponding amplitudes $d_k$ and can also be detected by analyzing
the shifted decimated signal, {\it i.e.\/}, signal points $c_1,\dots,c_{2K}$
instead of $c_0,\dots,c_{2K-1}$.
The true frequencies usually agree to very high precision, while spurious
frequencies show by orders of magnitude larger deviations.

\section{Results and discussion}
\label{results:sec}
In this section we want to demonstrate the efficiency and accuracy of
the method introduced above by way of two examples:
the three-disk repeller as an open physical system and the zeros of the
Riemann zeta function as a mathematical model for periodic orbit
quantization of bound chaotic systems.
Both systems have previously been investigated by means of
FD \cite{Mai97a,Mai98a,Mai99c}, which allows us
to make a direct comparison of the results.
The three-disk scattering system has also served as a prototype for 
the development and 
application of cycle expansion techniques \cite{Cvi89,Eck93,Wir99},
and we will briefly discuss the differences between harmonic inversion and 
cycle expansion.

\subsection{The three-disk repeller}
\label{3disk:subsec}
As the first example, we consider a billiard system consisting of three
identical hard disks with unit radii, $R=1$, displaced from each other
by the same distance $d$.
This simple, albeit nontrivial, scattering system has already been 
used as a model within the cycle expansion method 
\cite{Cvi89,Eck93,Wir99} and periodic orbit quantization by
harmonic inversion \cite{Mai97a,Mai98a,Mai99c}.
We give therefore only a very brief introduction to the system and refer
the reader to the literature for details.
The three-disk scattering system is invariant under the symmetry operations
of the group $C_{3v}$, {\it i.e.\/}, three reflections at symmetry lines 
and two rotations by $2\pi/3$ and $4\pi/3$.
Resonances belong to one of the three irreducible subspaces $A_1$, $A_2$, 
and $E$ \cite{Cvi93a}.
As in most previous work we concentrate on the resonances of the 
subspace $A_1$ for the three-disk repeller with $R=1$ and $d=6$.

\begin{figure*}
\vspace{18.5cm}
\includegraphics{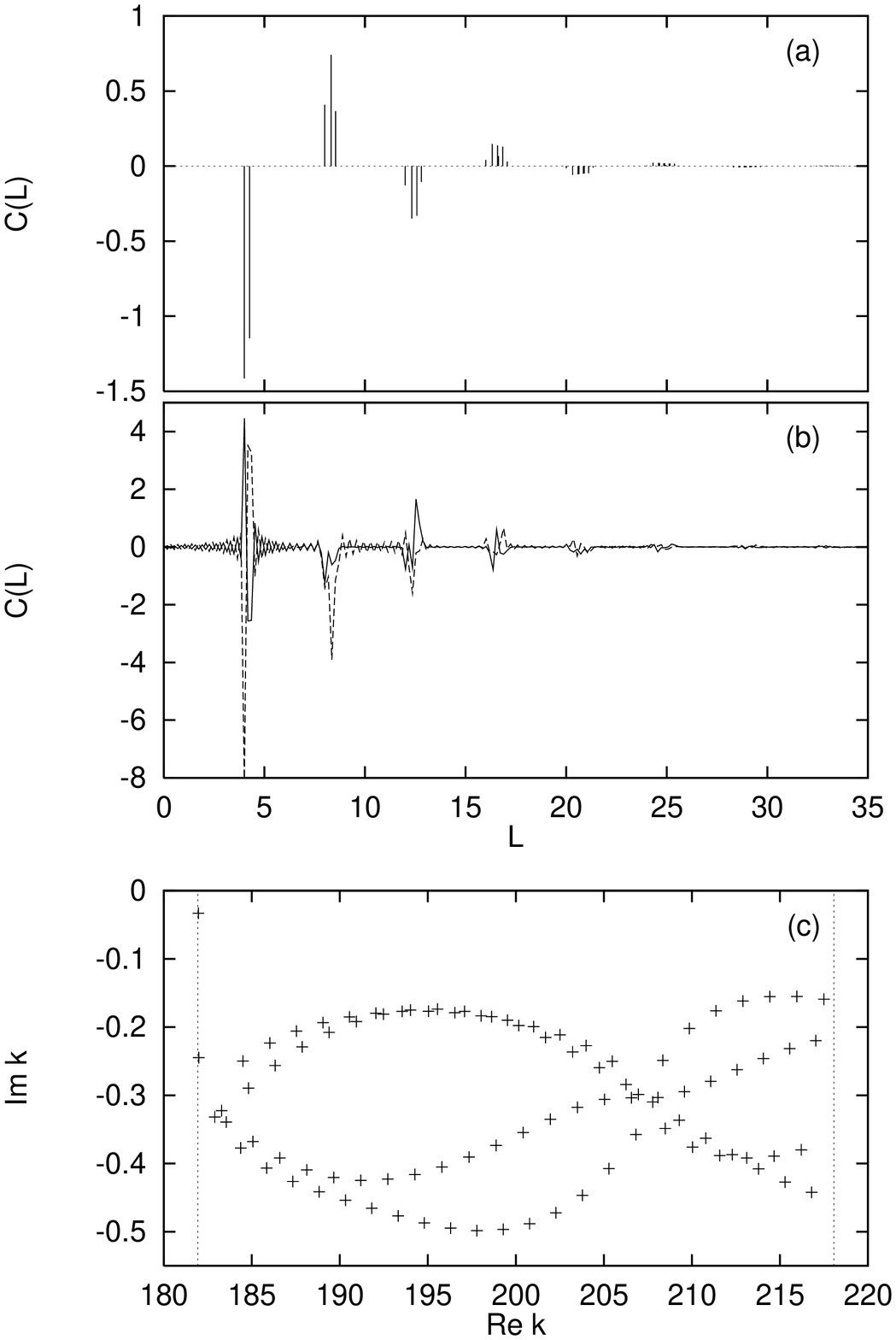}
\caption{(a) Periodic orbit recurrence signal for the three-disk scattering 
system with $R=1$, $d=6$ without filtering.
The signal in the region $L\le 35$ consists of 93 non-equidistant 
periodic orbit contributions (including multiple repetitions).
(b) Same as (a) filtered with frequency window $w\in [182, 218]$.
The decimated signal consists of 201 equidistant data points with 
$\Delta L=0.175$.
The solid and dashed lines are the real and imaginary part of $C(L)$,
respectively.
(c) Semiclassical resonances obtained by harmonic inversion of the
decimated signal $C(L)$ in (b).
The dotted lines mark the borders of the frequency window.
}
\label{Fig:1}
\end{figure*}

In billiards, which are scaling systems, the shape of the periodic orbits
does not depend on the energy $E$, and the classical action is given by 
the length $L$ of the orbit, {\it i.e.\/}, 
$S_{\rm po}=ws_{\rm po}=\hbar kL_{\rm po}$ (see eq.~\eref{S_po}), where 
$w=k=|{\bf k}|=\sqrt{2ME}/\hbar$ is the absolute value of the wave vector 
to be quantized.
Setting $\hbar=1$, we use $s_{\rm po}=L_{\rm po}$ in what follows.
In \fref{Fig:1}a we present the periodic orbit signal $C^{\rm sc}(L)$
for the three-disk repeller in the region $0\le L\le L_{\rm max}=35$.
The signal is given as a periodic orbit sum of delta functions
$\delta(L-L_{\rm po})$ weighted with the periodic orbit amplitudes
${\cal A}_{\rm po}$ (see eq.~\eref{C_sc}).
The groups with oscillating sign belong to periodic orbits with adjacent
cycle lengths.
Signals of this type have been analyzed (after convolution with a narrow
Gaussian function, see eq.~\eref{C_sc_sigma}) by FD
in refs.~\cite{Mai97a,Mai98a,Mai99a,Mai99b,Mai99c}.
We now illustrate harmonic inversion of band-limited decimated periodic orbit 
signals by DLP, DPA and DSD.

In a first step, a band-limited decimated periodic orbit signal is constructed
as described in \sref{analyt_dec:sec}.
As an example we choose $K=100$ as the rank of the nonlinear set of 
equations \eref{C_bld}, and $k_0=200$ as the centre of the frequency window.
The width of the frequency window is given by 
$\Delta k=2\pi K/L_{\rm max}=200\pi/35\approx 18.0$.
The step width of the decimated signal is $\tau=\Delta L=L_{\rm max}/2K=0.175$.
The band-limited decimated periodic orbit signal points 
$c_n=C^{\rm sc}_{\rm bl}(L=n\Delta L)$, with $n=0,\dots,2K$ are 
calculated with the help of eq.~\eref{C_sc_bl} and presented in \fref{Fig:1}b.
The solid and dashed lines are the real and imaginary parts of 
$C^{\rm sc}_{\rm bl}(n\Delta L)$, respectively.
The modulations with spacings $\pi/\Delta k$ result from the superposition
of the sinc-like functions in eq.~\eref{C_sc_bl}.

The band-limited decimated periodic orbit signal 
$C^{\rm sc}_{\rm bl}(n\Delta L)$
can now be analyzed, in a second step, with one of the harmonic inversion
techniques introduced in \sref{hi:sec}, {\it viz.\/}\ DLP, DPA or DSD.
The resonances obtained by DLP are presented as plus symbols in \fref{Fig:1}c.
The dotted lines at ${\rm Re}~k=182$ and ${\rm Re}~k=218$ show the borders
of the frequency window.
The two symbols very close to the border on the left indicate spurious 
resonances.

\begin{figure*}
\vspace{18.5cm}
\includegraphics{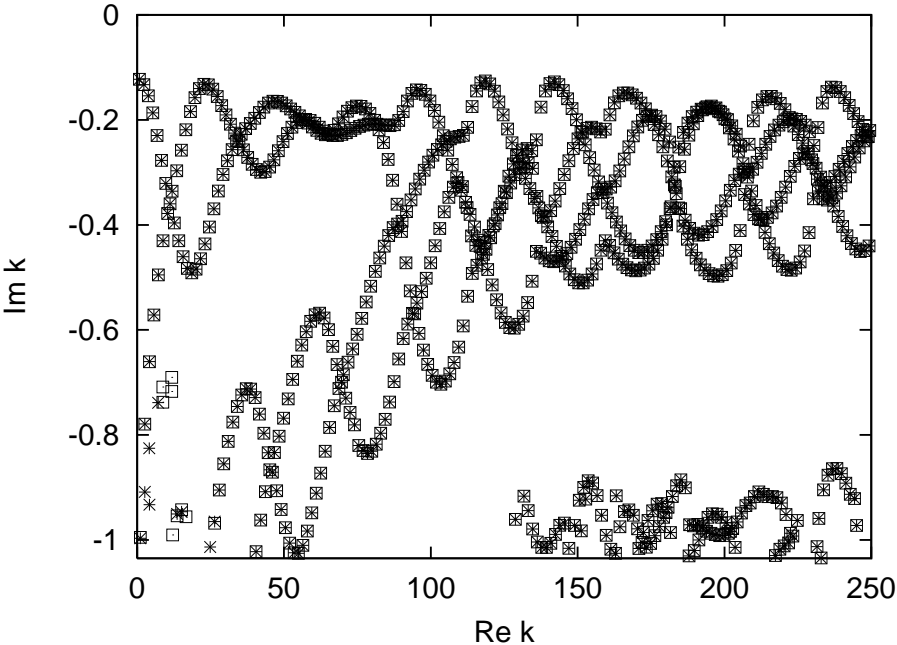}
\caption{Semiclassical resonances for the three-disk scattering system 
($A_1$ subspace) with 
$R=1$, $d=6$ obtained by harmonic inversion via
Decimated Linear Predictor (plus symbols),
Decimated Pad\'e Approximant (crosses), and
 Decimated Signal Diagonalization (squares)
of the analytically decimated periodic orbit signal.
}
\label{Fig:2}
\end{figure*}

A long range spectrum can be obtained by choosing several values $w_0$ for
the centre of the frequency window in such a way that the windows slightly
overlap.
\Fref{Fig:2} presents the semiclassical resonances for the three-disk repeller
in the range $0\le {\rm Re}~k\le 250$.
The spectrum has been obtained by harmonic inversion of decimated periodic 
orbit signals similar to that in \fref{Fig:1}b but with an increased 
signal length, $L_{\rm max}=55$.
The plus symbols, crosses, and squares denote the semiclassical resonances
obtained by DLP, DPA and DSD, respectively.
The resonances obtained by the three different harmonic inversion techniques
are in perfect agreement, with the exception of a few resonances in the region
${\rm Re}~k<25$.
In this region the matrices ${\bf U}$ and ${\bf S}$ in eq.~\eref{matels} are
rather ill-conditioned, and the few discrepancies can therefore be explained 
as numerical artifacts.

The spectrum presented in \fref{Fig:2} was obtained previously in 
ref.~\cite{Mai98a} by application of FD \cite{Wal95,Man97a,Man97b}.
In \tref{table1} we compare the semiclassical eigenvalues $k$ and residues 
$d_k$ of selected resonances obtained by (a) FD and (b) harmonic inversion 
of band-limited decimated periodic orbit signals.
For non-degenerate resonances under study the residues should be $d_k=1$.
In \cite{Mai98a} the residues of several resonances deviate significantly 
from $d_k=1$ by more than $5\%$ 
(see the resonances marked by $(^a)$ in \tref{table1}).
With harmonic inversion of decimated signals the accuracy of the
residues is increased by several orders of magnitude
(see the resonances marked by $(^b)$ in \tref{table1}).
The semiclassical eigenvalues $k$ also reveal deviations between the
different numerical techniques.
The resonances obtained by harmonic inversion of band-limited decimated 
signals are in much better agreement with the results obtained by the cycle 
expansion method \cite{Wir99} than those obtained by FD
in ref.~\cite{Mai98a}.

Numerical values for the residues very close to $d_k=1$ indicate well 
converged semiclassical resonances, and this is the case for all resonances 
of the four bands closest to the real axis in \fref{Fig:2}.
Resonances with non-vanishing $d_k$ have also been obtained in the region 
${\rm Re}~k>120$, ${\rm Im}~k<-0.8$ (see \fref{Fig:2}).
These resonances, although not fully converged, are in qualitative 
agreement with exact quantum calculations \cite{Wir99}.
It is important to note that the different techniques for harmonic inversion 
of decimated signals, {\it viz.\/}\ DLP, DPA and DSD, yield the same results,
even for those resonances which are not fully converged.
This illustrates the mathematical equivalence of the three methods as 
explained in \sref{hi:sec}.

\subsection{Harmonic inversion vs.\ cycle expansion}
The three-disk scattering system discussed in \sref{3disk:subsec} has
purely hyperbolic classical dynamics and has 
been used extensively as the prototype model within the
cycle expansion techniques \cite{Cvi89,Eck93,Wir99}.
As has been shown by Voros \cite{Vor88}, Gutzwiller's trace formula for
unstable periodic orbits can be recast in the form of
an infinite and non-convergent Euler product over all periodic orbits.
When the periodic orbits obey a symbolic dynamics the semiclassical 
eigenenergies or resonances can be obtained as the zeros of the cycle
expanded Gutzwiller-Voros zeta function.
Unfortunately, the convergence of the cycle expansion is restricted, due to
poles of the Gutzwiller-Voros zeta function \cite{Eck93}.
The domain of analyticity of semiclassical zeta functions can be extended 
\cite{Cvi93b,Cvi93c} resulting in
the ``quasiclassical zeta function'' \cite{Cvi93c,Wir99}, which is an entire 
function for the three-disk repeller.
This approach allows one to calculate semiclassical resonances in critical 
regions where the Gutzwiller-Voros zeta function does not converge, at the 
cost, however, of many extra spurious resonances and with the rate of 
convergence slowed down tremendously \cite{Wir99}.

With the limited numerical accuracy of harmonic inversion by
FD applied in ref.~\cite{Mai98a}, the semiclassical 
resonances of the three-disk repeller in the region ${\rm Im}~k<-0.6$ 
were somewhat unreliable.
The improved accuracy of the analysis of band-limited decimated periodic
orbit signals introduced in the present paper now allows us to compare 
the two semiclassical quantization techniques, {\it viz.\/}\ harmonic 
inversion and cycle expansion methods, even for resonances deep in the 
complex plane.
We will demonstrate that the harmonic inversion method provides semiclassical
resonances in energy regions where the cycle expansion of the Gutzwiller-Voros
zeta function does not converge.

\begin{figure*}
\vspace{18.5cm}
\includegraphics{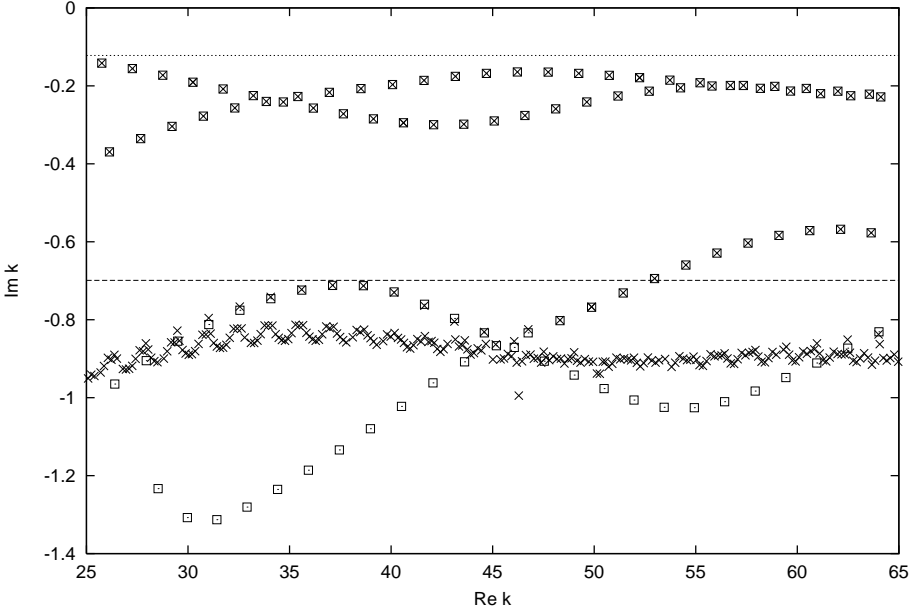}
\caption{
Semiclassical resonances ($A_1$ subspace) for the three-disk scattering  
system with $R=1$, $d=6$. 
Squares: Harmonic inversion of the decimated semiclassical recurrence signal;
Crosses: Cycle expansion of the Gutzwiller-Voros zeta function
\protect\cite{Wir99}.
The dotted and dashed lines mark the borderline for absolute convergence
of Gutzwiller's trace formula $({\rm Im}~k=-0.121\,557)$ and the
Gutzwiller-Voros zeta function $({\rm Im}~k=-0.699\,110)$, respectively.
The harmonic inversion method converges deeper in the complex plane than 
the Gutzwiller-Voros zeta function.
}
\label{Fig:3}
\end{figure*}

In \fref{Fig:3} we present a part of the semiclassical resonance spectrum
of \fref{Fig:2} in the region $25 \le {\rm Re}~k \le 65$.
The squares and crosses  label the semiclassical resonances obtained by 
harmonic inversion of the decimated semiclassical periodic orbit signal 
and cycle expansion of the Gutzwiller-Voros zeta function \cite{Wir99}, 
respectively.
The dotted line in \fref{Fig:3} indicates the borderline, 
${\rm Im}~k=-0.121\,557$ \cite{Cvi89}, which separates the domain of 
absolute convergence of Gutzwiller's trace formula from the region where 
analytic continuation is necessary.
For the two resonance bands slightly below this border the results of both
semiclassical quantization methods are in perfect agreement.
The dashed line in \fref{Fig:3} marks the abscissa of absolute convergence 
for the Gutzwiller-Voros zeta function at ${\rm Im}~k=-0.699\,110$ 
\cite{Cvi93b}.
The Gutzwiller-Voros zeta function provides several spurious resonances
which accumulate at ${\rm Im}~k\approx -0.9$, {\it i.e.\/}, slightly below 
the borderline of absolute convergence (see the crosses in \fref{Fig:3}).
The resonances in the region ${\rm Im}~k<-0.9$, 
especially those belonging to the fourth band,
are not described by the Gutzwiller-Voros zeta function but are obtained
by the harmonic inversion method (see the squares in \fref{Fig:3}).

\subsection{Zeros of the Riemann zeta function}
As the second example to demonstrate the numerical accuracy of harmonic
inversion of band-limited decimated periodic orbit signals we investigate
the Riemann zeta function which is a mathematical model for
periodic orbit quantization.
Here we only briefly explain the idea of this model and refer the reader
to the literature \cite{Mai98a,Mai99c} for details.

The hypothesis of Riemann is that all the nontrivial zeros of the analytic
continuation of the function
\begin{equation}
   \zeta(z)
 = \sum_{n=1}^\infty n^{-z}
 = \prod_p \left(1-p^{-z}\right)^{-1} \; , 
   \quad ({\rm Re}~ z>1,~ p: {\rm primes})
\label{zeta_def}
\end{equation}
have real part $1\over 2$, so that the values $w=w_k$, defined by
\begin{equation}
 \zeta\left(\case{1}{2}-iw_k\right) = 0 \; ,
\end{equation}
are all real or purely imaginary \cite{Edw74,Tit86}.
The parameters $w_k$ can be obtained as the poles of the function
\begin{equation}
 g(w) = \sum_p \sum_{m=1}^\infty {\cal A}_{pm} e^{iws_{pm}} \; ,
\label{g_nc}
\end{equation}
where
\begin{eqnarray}
\label{Apm}
 {\cal A}_{pm}&=& i{\log(p)\over p^{m/2}} \; , \\
\label{Spm}
 s_{pm}&=&m\log(p) \; ,
\end{eqnarray}
with $p$ indicating the prime numbers.
As was already pointed out by Berry \cite{Ber86}, eq.~\eref{g_nc} has
the same mathematical form as Gutzwiller's trace formula with the primes
interpreted as the primitive periodic orbits, ${\cal A}_{pm}$ and 
$s_{pm}$ the ``amplitudes'' and ``classical actions'' of the periodic orbit
contributions, and $m$ formally counting the ``repetitions'' of orbits.
Equation~\eref{g_nc} converges only for ${\rm Im}~w>{1\over 2}$ and analytic
continuation is necessary to extract the poles of $g(w)$, {\it i.e.\/}, the
Riemann zeros.
The advantage of studying the zeta function instead of a ``real'' physical 
bound system is that no extensive periodic orbit search is necessary for 
the calculation of Riemann zeros, as the only input data are just prime 
numbers.
Harmonic inversion can be applied to adjust the Fourier transform of
eq.~\eref{g_nc}, {\it i.e.\/},
\begin{equation}
  C(s) = \sum_p \sum_{m=1}^\infty {\cal A}_{pm} \delta(s-s_{pm}) \; ,
\label{C_nc}
\end{equation}
to the functional form
\begin{equation}
   C_{\rm ex}(s)
 = {1\over 2\pi} \int_{-\infty}^{+\infty} \sum_k 
    {d_k \over w-w_k+i\epsilon} e^{-isw} dw
 = -i \sum_k d_k e^{-iw_ks} \; ,
\label{C_ex}
\end{equation}
where the $w_k$ are the Riemann zeros and the residues $d_k$ 
have been introduced as adjustable parameters which here should all 
be equal to 1.

In ref.~\cite{Mai98a} about 2600 Riemann zeros to about 12 digit precision
have been obtained by harmonic inversion of the signal \eref{C_nc} with 
$s_{\rm max}=\log(10^6)\approx 13.82$ using FD.
However, the numerical residues agree with $d_k=1$ only to about a
5 or 6 digit precision.
With harmonic inversion of band-limited decimated signals the accuracy
of both the Riemann zeros $w_k$ and the multiplicities $d_k$ is improved
by several orders of magnitude.
In \tref{table2} we compare selected values obtained by (a) 
FD \cite{Mai98a} and (b) DLP.
The increase in accuracy can easily be seen, in particular for the
imaginary parts, ${\rm Im}~w_k$ and ${\rm Im}~d_k$, which should both be
equal to zero.
The same improvement in accuracy is also achieved by application of 
DPA and DSD.

The precise calculation of parameters $d_k=1$ for the residues of
the Riemann zeros does not seem to be of great interest.
However, it should be noted that the multiplicities may be greater than
one, {\it e.g.\/}, for some eigenvalues of integrable systems such as the
circle billiard \cite{Mai99c,Mai99e} where states with angular momentum 
quantum number $m\ne 0$ are twofold degenerate.
As we have checked the techniques presented in this paper indeed yield
the correct multiplicities to very high precision.
The $d_k$'s have also nontrivial values when used, {\it e.g.\/}, for the 
semiclassical calculation of diagonal matrix elements \cite{Mai99e} and 
non-diagonal transition strengths \cite{Mai99a} in dynamical systems.

\section{Conclusion}
\label{concl:sec}
We have introduced three methods for semiclassical periodic orbit quantization,
{\it viz.\/}\ Decimated Linear Predictor (DLP), Decimated Pad\'e Approximant 
(DPA), and Decimated  Signal Diagonalization (DSD) for the harmonic inversion 
of band-limited decimated periodic orbit signals.
The characteristic feature of these methods is the strict separation of the 
two steps, {\it viz.\/}\ the analytical filtering of the periodic orbit signal 
and the numerical harmonic inversion of the band-limited decimated signal.
The separation of these two steps and the handling of small amounts of data
compared to other ``black box'' type signal processing techniques enables an 
easier and deeper understanding of the semiclassical quantization method.
Furthermore, applications to the three-disk repeller and the Riemann zeta
function demonstrate that the new methods provide numerically more accurate
results than previous applications of filter-diagonalization (FD).
A detailed comparison of various semiclassical quantization methods reveals 
that quantization by harmonic inversion of the band-limited decimated 
periodic orbit signal can even be applied 
in energy regions where the cycle expansion of the Gutzwiller-Voros zeta 
function does not converge.

The methods introduced in this paper can be applied to the periodic orbit 
quantization of systems with both chaotic and regular classical dynamics, 
when the periodic orbit signal is calculated with Gutzwiller's trace 
formula \cite{Gut67,Gut90} for isolated orbits and the Berry-Tabor 
formula \cite{Ber76} for orbits on rational tori, respectively.
More generally, any signal given as a sum of $\delta$ functions can be
filtered analytically and analyzed using the methods described in sections 
\ref{analyt_dec:sec} and \ref{hi:sec}.
For example, the technique can also be applied to the harmonic inversion 
of the density of states $\varrho(E)=\sum_n\delta(E-E_n)$ of quantum 
spectra to extract information about the underlying classical dynamics 
\cite{Mai97b,Mai99c}.

It is to be noted that all the signal processing techniques mentioned in
this paper lend themselves to a formulation where non-diagonal responses
appear in the frequency domain and their complementary cross-correlation-type
signals appear in the time or action domain \cite{Nar97,Mai99e}.
This is important as such methods allow for the use of shorter signals
and hence, in the context of this paper, fewer periodic orbits.
The need to find large numbers of periodic orbits may limit the practical 
utility of these methods and so any attempt to overcome this problem is 
worth investigating.
Cross-correlation methods also sample better and yield improved results for 
poles (resonances) that lie deep in the complex plane than do straight 
correlation-based signal processing techniques.

\ack 
%
We thank A.\ Wirzba for supplying numerical data on the three-disk system.
This work was supported in part by the Deutsche Forschungsgemeinschaft
(Grant number WU130/12-1) and the National Science Foundation
(Grant number PHY-9802534).

\section*{References}

\newpage

\newpage
%
\begin{table}[t]
\caption{\label{table1}
Semiclassical resonances and multiplicities for the three-disk scattering 
problem ($A_1$ subspace) with $R=1$, $d=6$. 
($^a$) Filter-diagonalization method (FD) \protect\cite{Mai98a};
($^b$) Decimated Linear Predictor (DLP).
}

\bigskip
\begin{center}
\begin{tabular}[t]{rr|rr}
  \multicolumn{1}{c}{${\rm Re}~k$} &
  \multicolumn{1}{c|}{${\rm Im}~k$} &
  \multicolumn{1}{c}{${\rm Re}~d_k$} &
  \multicolumn{1}{c}{${\rm Im}~d_k$} \\ \hline
$^a$  126.16812780 &   -0.21726568 &    0.99997532 &    0.00000523 \\
$^b$  126.16812767 &   -0.21726638 &    0.99999999 &    0.00000002 \\[1ex]
$^a$  126.57000032 &   -0.30717994 &    0.99969830 &   -0.00028769 \\
$^b$  126.57000899 &   -0.30718955 &    1.00000078 &   -0.00000009 \\[1ex]
$^a$  126.89863330 &   -0.61058335 &    1.24854908 &   -0.16290432 \\
$^b$  126.90658296 &   -0.59570469 &    1.00062326 &   -0.00390071 \\[1ex]
$^a$  127.21759681 &   -0.32010287 &    1.00042752 &    0.00045232 \\
$^b$  127.21758249 &   -0.32008945 &    0.99999992 &    0.00000075 \\[1ex]
$^a$  127.68308651 &   -0.24341398 &    0.99993610 &   -0.00000236 \\
$^b$  127.68308662 &   -0.24341588 &    1.00000001 &    0.00000005 \\[1ex]
$^a$  128.12116088 &   -0.28389637 &    1.00010175 &   -0.00022229 \\
$^b$  128.12116753 &   -0.28389340 &    1.00000025 &    0.00000034 \\[1ex]
$^a$  128.41137217 &   -0.61577414 &    1.32422538 &   -0.11565068 \\
$^b$  128.41689182 &   -0.59664661 &    1.00031395 &   -0.00566184 \\[1ex]
$^a$  128.70334065 &   -0.30442655 &    1.00039570 &    0.00011310 \\
$^b$  128.70333742 &   -0.30441448 &    1.00000039 &    0.00000030 \\[1ex]
$^a$  129.19732946 &   -0.26788859 &    0.99987656 &   -0.00004493 \\
$^b$  129.19733098 &   -0.26789240 &    1.00000003 &    0.00000011 \\[1ex]
$^a$  129.67319699 &   -0.26717842 &    1.00017817 &    0.00002664 \\
$^b$  129.67319613 &   -0.26717327 &    0.99999982 &    0.00000017 \\[1ex]
$^a$  129.92927207 &   -0.60918315 &    1.33322988 &   -0.02170382 \\
$^b$  129.93018579 &   -0.58921656 &    1.00070097 &   -0.00611400 \\[1ex]
$^a$  130.18796223 &   -0.29223540 &    1.00028313 &   -0.00012743 \\
$^b$  130.18796644 &   -0.29222731 &    1.00000035 &   -0.00000011 \\[1ex]
$^a$  130.71098079 &   -0.29045241 &    0.99983213 &   -0.00012609 \\
$^b$  130.71098501 &   -0.29045760 &    1.00000005 &    0.00000021 \\[1ex]
$^a$  131.22717821 &   -0.25736473 &    1.00000071 &    0.00017324 \\
$^b$  131.22717326 &   -0.25736500 &    0.99999989 &   -0.00000015 \\[1ex]
$^a$  131.44889208 &   -0.59054385 &    1.26610320 &    0.06973001 \\
$^b$  131.44497889 &   -0.57352843 &    1.00129044 &   -0.00468689
\end{tabular}
\end{center}
\end{table}

\begin{table}[c]
\caption{\label{table2}
Nontrivial zeros $w_k$ and multiplicities $d_k$ for the Riemann zeta 
function.
($^a$) Filter-diagonalization method (FD) \protect\cite{Mai98a};
($^b$) Decimated Linear Predictor (DLP).
}

\bigskip
\begin{center}
\begin{tabular}[t]{r|rr|rr}
  \multicolumn{1}{c|}{$k$} &
  \multicolumn{1}{c}{${\rm Re}~w_k$} &
  \multicolumn{1}{c|}{${\rm Im}~w_k$} &
  \multicolumn{1}{c}{${\rm Re}~d_k$} &
  \multicolumn{1}{c}{${\rm Im}~d_k$} \\ \hline
$^a$    1 &   14.13472514 &  4.05E-12 &  1.00000011 & -5.07E-08 \\
$^b$    1 &   14.13472514 & -7.43E-15 &  1.00000000 &  2.63E-13 \\[1ex]
$^a$    2 &   21.02203964 & -2.23E-12 &  1.00000014 &  1.62E-07 \\
$^b$    2 &   21.02203964 &  2.48E-14 &  1.00000000 &  3.71E-13 \\[1ex]
$^a$    3 &   25.01085758 &  1.66E-11 &  0.99999975 & -2.64E-07 \\
$^b$    3 &   25.01085758 & -4.70E-14 &  1.00000000 & -7.04E-14 \\[1ex]
$^a$    4 &   30.42487613 & -6.88E-11 &  0.99999981 & -1.65E-07 \\
$^b$    4 &   30.42487613 &  2.82E-13 &  1.00000000 &  5.58E-13 \\[1ex]
$^a$    5 &   32.93506159 &  7.62E-11 &  1.00000020 &  5.94E-08 \\
$^b$    5 &   32.93506159 &  3.47E-14 &  1.00000000 &  1.14E-13 \\
$\cdots$ &  $\cdots$ &  $\cdots$ &  $\cdots$ &  $\cdots$ \\
$^a$ 2561 & 3093.18544571 & -2.33E-09 &  1.00000168 & -1.50E-07 \\
$^b$ 2561 & 3093.18544572 & -2.77E-13 &  1.00000000 &  1.08E-11 \\[1ex]
$^a$ 2562 & 3094.83306842 &  2.07E-08 &  0.99999647 &  2.63E-06 \\
$^b$ 2562 & 3094.83306843 & -8.77E-12 &  1.00000000 & -5.20E-11 \\[1ex]
$^a$ 2563 & 3095.13203122 & -1.79E-08 &  1.00000459 &  1.70E-06 \\
$^b$ 2563 & 3095.13203124 &  3.70E-12 &  1.00000000 &  2.86E-11 \\[1ex]
$^a$ 2564 & 3096.51548551 &  5.15E-09 &  0.99999868 &  2.74E-06 \\
$^b$ 2564 & 3096.51548551 & -9.16E-13 &  1.00000000 & -5.44E-12 \\[1ex]
$^a$ 2565 & 3097.34260655 &  7.75E-09 &  0.99999918 &  5.12E-06 \\
$^b$ 2565 & 3097.34260653 & -1.98E-12 &  1.00000000 & -1.56E-11
\end{tabular}
\end{center}
\end{table}

\end{document}